%Paper: astro-ph/9406073
%From: yzqian@ben.npl.washington.edu
%Date: Wed, 29 Jun 94 12:00:58 -0700

\magnification=1200
\baselineskip=24pt plus 2pt
\centerline{\bf Neutrino-Neutrino Scattering and Matter-Enhanced}
\centerline{\bf  Neutrino Flavor Transformation in Supernovae} 
\centerline{{\sl Yong-Zhong Qian} and {\sl George M.  
Fuller}\footnote{$^1$}{Permanent address: Department of Physics,  
University of California, San Diego, La Jolla, CA 92093-0319.}}
\centerline{Institute for Nuclear Theory, HN-12}
\centerline{University of Washington, Seattle, WA 98195}
\centerline{\sl ABSTRACT}
We examine matter-enhanced neutrino flavor transformation  
($\nu_{\tau(\mu)}\rightleftharpoons\nu_e$) in the region above the  
neutrino sphere in Type II supernovae. Our treatment explicitly  
includes contributions to the neutrino-propagation Hamiltonian from  
neutrino-neutrino forward scattering. A proper inclusion of these  
contributions shows that they have a completely negligible effect on  
the range of $\nu_e$-$\nu_{\tau(\mu)}$ vacuum mass-squared  
difference, $\delta m^2$, and vacuum mixing angle, $\theta$, or  
equivalently $\sin^22\theta$, required for enhanced supernova shock  
re-heating. When neutrino background effects are included, we find  
that $r$-process nucleosynthesis from neutrino-heated supernova  
ejecta remains a sensitive probe of the mixing between a light  
$\nu_e$ and a $\nu_{\tau(\mu)}$ with a cosmologically significant  
mass. Neutrino-neutrino scattering contributions are found to have a  
generally small effect on the $(\delta m^2,\ \sin^22\theta)$  
parameter region probed by $r$-process nucleosynthesis. We point out
that the nonlinear effects of the neutrino background extend the range
of sensitivity of $r$-process nucleosynthesis to smaller values of $\delta m^2$.

\vfill
\eject
\centerline{\bf I. Introduction}
\vskip .2in
In this paper we investigate the problem of matter-enhanced neutrino  
flavor transformation in the region above the neutrino sphere in Type
II supernovae. In particular, we examine the role of contributions to  
the neutrino-propagation Hamiltonian from neutrino-neutrino forward  
scattering. A general framework for treating these contributions in the
context of the Mikeheyev-Smirnov-Wolfenstein (MSW) neutrino flavor
transformation process has been given in Ref. [1] (see Ref. [2] for a
numerical study of the case of a pure neutrino gas). Although the role
of neutrino-neutrino scattering in the problem of matter-enhanced
neutrino flavor conversion in supernovae has been treated previously
[3, 4], the present paper gives the first complete treatment utilizing
the scheme of Ref. [1]. 

Recent studies have examined MSW tranformation of $\nu_\tau$ or  
$\nu_\mu$ into $\nu_e$ in the region above the neutrino sphere in the  
post-core-bounce supernova environment [5, 6]. These studies suggest  
that if $\nu_\tau$ or $\nu_\mu$ has a mass in the cosmologically  
interesting range of 1--100 eV, then matter-enhanced transformation  
to $\nu_e$ will be possible in this region. Such transformation can  
result in significant effects on supernova dynamics and/or  
nucleosynthesis.

If we define, for example, $|\nu_e\rangle$ and $|\nu_\tau\rangle$ to  
be flavor eigenstates of $\nu_e$ and $\nu_\tau$, and $|\nu_1\rangle$  
and $|\nu_2\rangle$ to be the associated mass eigenstates, then the  
vacuum mixing angle, $\theta$, is defined through
$$\eqalignno{|\nu_e\rangle&=\cos\theta|\nu_1\rangle+\sin\theta|\nu_2\rangle,&(1a)\cr
|\nu_\tau\rangle&=-\sin\theta|\nu_1\rangle+\cos\theta|\nu_2\rangle.&(1b)
\cr}$$
Ref. [5] shows that $\nu_{\tau(\mu)}\rightleftharpoons\nu_e$ mixing  
with $\sin^22\theta\ge10^{-7}$ in the region above the neutrino  
sphere at a few hundred milliseconds after the bounce of the core can  
result in a (30--60)\% increase in the supernova shock energy. Ref.  
[6] shows that the heavy element nucleosynthesis from the hot bubble  
region is sensitive to $\nu_{\tau(\mu)}\rightleftharpoons\nu_e$  
mixing at a level of $\sin^22\theta\sim10^{-5}$. This hot bubble region
forms above the neutrino sphere $\sim 3$ seconds after  
core bounce. These effects are sensitive to mixing angles far smaller
than those which can be probed in laboratory  
experiments. These supernova effects ultimately may represent our  
most sensitive probe of putative neutrino dark matter.

However, studies [5] and [6] neglected the off-diagonal contributions  
of neutrino-neutrino scattering to the flavor-basis  
neutrino-propagation Hamiltonian. In what follows, we present a  
detailed study of neutrino flavor transformation in the  
post-core-bounce supernova environment. Our calculations include all  
effects of the neutrino background.
We have adopted the overall principles and techniques of Ref. [1] in  
our treatment of neutrino-neutrino and neutrino-electron scattering  
contributions to the neutrino-propagation Hamiltonian. We find that  
neutrino background contributions have a negligible effect on the  
range of $\nu_e$-$\nu_{\tau(\mu)}$ vacuum mass-squared difference,  
$\delta m^2$, and vacuum mixing angle, $\theta$ (or $\sin^22\theta$),  
required for enhanced supernova shock re-heating. A proper treatment  
of the ensemble average over the neutrino background shows that  
$r$-process nucleosynthesis from neutrino-heated supernova ejecta  
remains a sensitive probe of the mixing between a light $\nu_e$ and a  
$\nu_\tau$ (or $\nu_\mu$) with a cosmologically significant mass  
($m_{\nu_{\tau(\mu)}}\approx 1$--100 eV).

In Sec. II we discuss a general framework for treating neutrino  
flavor transformation in the supernova environment. In Sec. III we  
compute neutrino flavor transformation probabilities as functions of  
$\delta m^2$ and $\sin^22\theta$ relevant for the shock re-heating  
and hot bubble/$r$-process nucleosynthesis epochs of the supernova.  
We give conclusions in Sect. IV.
\vskip .2in
\centerline{\bf II. The Neutrino-Propagation Hamiltonian in  
Supernovae}
\vskip .2in
The general problem of the time evolution of the full density matrix  
for an ensemble of three flavors of neutrinos and antineutrinos with  
an electron/positron background and a nucleon background is a  
daunting one. Several formal approaches to this problem have been  
made (cf. Ref. [1] and references therein). In the present paper, we  
shall only summarize the salient features of this previous work and  
taylor our subsequent discussion to the particular problem of  
neutrino propagation and flavor transformation in the region of the  
supernova environment above the neutrino sphere. Considerable  
simplification of the problem can be realized in this case.

The general time evolution of the neutrino density matrix $\rho$ can  
be summarized as
$$i\dot\rho=[H,\rho],\eqno(2)$$
where $\rho=\sum_{ij}\rho_{ij}|i\rangle\langle j|$,  
$\dot\rho=d\rho/dt$, and $i$ and $j$ refer to {\it all} neutrino  
quantum numbers including momentum (energy), flavor, helicity, charge  
conjugation eigenvalue, etc. In Eq. (2), $H$ is the full  
Schr\"odinger picture Hamiltonian including all neutrino self  
interactions as well as interactions with the $e^\pm$ and nucleon  
backgrounds.

Without loss of generality we can follow a particular momentum  
component of Eq. (2) (cf. Ref. [1]), or equivalently, the associated  
Schr\"odinger equation for the time evolution of neutrino field  
amplitudes for a given momentum. The Hamiltonian operator in this  
case would have the dimensionality of the density matrix for the  
single momentum state (e.g., 12$\times12$ for three Dirac neutrino  
flavors, since each neutrino state has either right-handed or  
left-handed helicity, and is either a neutrino or an antineutrino).

We argue that further simplification  of this problem can be made  
through approximations motivated by the {\it particular} distribution  
functions for $\nu_e$, $\bar\nu_e$, $\nu_\mu$, $\bar\nu_\mu$,  
$\nu_\tau$, and $\bar\nu_\tau$ which obtain in the region above the  
neutrino sphere in the post-core-bounce epoch of Type II supernovae.  
Since the distribution functions for $\nu_\mu$, $\bar\nu_\mu$,  
$\nu_\tau$, and $\bar\nu_\tau$ are all expected to be essentially  
identical, mixings between neutrinos in this sector will have no  
effect on any aspect of supernova physics. In other words, we need  
only consider mixings between $\nu_e$ and either $\nu_\mu$ or  
$\nu_\tau$. If, as seems likely, the vacuum mass heirachy for  
neutrinos satisfies $m_{\nu_{\tau(\mu)}}>m_{\nu_e}$, then we need  
only consider matter-enhanced mixing among neutrinos, as antineutrino  
mixing is supressed by matter effects. 

The masses $m_{\nu_{\tau(\mu)}}\approx 1$--100 eV of interest in the  
post-core-bounce supernova environment are very small compared to the  
typical neutrino energies (average neutrino energy $\langle  
E_\nu\rangle$ is about or greater than 10 MeV). In this case we can  
neglect the population of right-handed Dirac neutrinos and  
left-handed Dirac antineutrinos produced by scattering processes.  
This is because helicity-flipping rates are proportional to  
$(m_\nu/E_\nu)^2$.

Taking advantage of these features allows us to reduce the  
dimensionality of the Hamiltonian in Eq. (2) to $2\times2$ for the  
Dirac neutrino case. If neutrinos are Majorana particles, then we  
have only left-handed neutrinos and right-handed antineutrinos, and  
again the Hamiltonian of interest is $2\times2$.

In any case, the neutrinos of interest in supernovae will be  
extremely relativistic, so that we can approximate the neutrino  
energy as $E_\nu=\sqrt{p^2+m^2}\approx p+m^2/2p$. The first term in  
this expression, $p$, the momentum, just gives an overall phase to  
the coherent propagating neutrino state and can be ignored without  
loss of generality. The second term, $m^2/2p$, is responsible for the  
relevant neutrino mixing behavior. The part of the Hamiltonian  
corresponding to the $m^2/2p$ term in vacuum, $H_{\rm v}$, can be  
written in the flavor basis (e.g., $|\nu_e\rangle$,  
$|\nu_\tau\rangle$) as
$$H_{\rm v}={\Delta\over 2}\pmatrix{-\cos2\theta&\sin2\theta\cr
\sin2\theta&\cos2\theta\cr},\eqno(3)$$
where $\theta$ is the vacuum mixing angle as in Eq. (1),  
$\Delta=\delta m^2/2E_\nu$, and $\delta m^2\equiv m_2^2-m_1^2$, with  
$m_1$ and $m_2$ the vacuum mass eigenvalues corresponding to the mass  
eigenstates $|\nu_1\rangle$ and $|\nu_2\rangle$, respectively.

In matter the relation between the flavor basis and the mass basis  
can be written as in Eq. (1), but with the vacuum mixing angle  
replaced by an appropriate matter mixing angle $\theta_n$. For  
illustrative purposes consider the case where the only contribution  
to the effective mass difference between neutrino flavors comes from  
charged-current exchange scattering on electrons. We take the net  
number density of electrons to be 
$$n_e\equiv n_{e^-}-n_{e^+},\eqno(4a)$$
where $n_{e^-}$ $(n_{e^+})$ is the total proper number density of  
negatrons (positrons). The electron fraction $Y_e$ is defined in  
terms of the total baryon rest mass density $\rho$ and Avogardro's  
number $N_{\rm A}$ by
$$n_e=\rho Y_eN_{\rm A}.\eqno(4b)$$
The contribution to the Hamiltonian from neutrino-electron exchange  
scattering is
$$A=\sqrt{2}G_{\rm F}n_e,\eqno(5)$$
where $G_{\rm F}$ is the Fermi constant. In Fig. 1 we show a generic  
Feynman graph for $\nu_e$-$e$ scattering. To obtain the result in Eq.  
(5) one must sum graphs for $\nu_e$-$e^-$ and $\nu_e$-$e^+$  
scattering over the appropriate $e^\pm$ distribution functions. In  
this case the neutrino-propagation Hamiltonian, $H_e$, can be written  
as
$$H_e={\Delta_{\rm eff}\over  
2}\pmatrix{-\cos2\theta_n&\sin2\theta_n\cr
\sin2\theta_n&\cos2\theta_n\cr}={1\over  
2}\pmatrix{-\Delta\cos2\theta+A&\Delta\sin2\theta\cr
\Delta\sin2\theta&\Delta\cos2\theta-A\cr},
\eqno(6)$$
where $\Delta_{\rm  
eff}=\sqrt{(\Delta\cos2\theta-A)^2+\Delta^2\sin^22\theta}$. In these  
expressions the matter mixing angle, $\theta_n$, is related to the  
vacuum mixing angle $\theta$ and the local net electron number  
density through
$$\sin2\theta_n={\Delta\sin2\theta\over
\sqrt{(\Delta\cos2\theta-A)^2+\Delta^2\sin^22\theta}},\eqno(7a)$$
$$\cos2\theta_n={\Delta\cos2\theta-A\over
\sqrt{(\Delta\cos2\theta-A)^2+\Delta^2\sin^22\theta}}.\eqno(7b)$$

The amplitudes for antineutrino-electron ($\bar\nu_e$-$e$) exchange  
scattering and neutrino-electron ($\nu_e$-$e$) exchange scattering  
have opposite signs. This implies that $\bar\nu_e$-$e$ exchange  
scattering gives a contribution $-A$ to the flavor-basis interaction  
Hamiltonian for $\bar\nu_e$. In this case the matter mixing angle for  
antineutrinos, $\bar\theta_n$, satisfies
$$\sin2\bar\theta_n={\Delta\sin2\theta\over
\sqrt{(\Delta\cos2\theta+A)^2+\Delta^2\sin^22\theta}},\eqno(8a)$$
$$\cos2\bar\theta_n={\Delta\cos2\theta+A
\over
\sqrt{(\Delta\cos2\theta+A)^2+\Delta^2\sin^22\theta}}.\eqno(8b)$$
We note that the vacuum mixing angles for the neutrino and  
antineutrino sectors are the same.

It is evident from Eqs. (7a \& b) that matter effects can give  
enhancement of flavor mixing in the neutrino sector. Mixing is  
maximal at a mass level crossing, or resonance, where  
$\Delta\cos2\theta=A$ [7]. On the other hand, Eqs. (8a \& b) show  
that matter effects give supression of flavor mixing in the  
antineutrino sector.

In the supernova environment, however, the neutrino background and  
the resultant  neutrino-neutrino forward exchange-scattering effects  
necessitate some modification of the above treatment of neutrino  
flavor transformation. In the region above the neutrino sphere in  
post-core-bounce Type II supernovae the neutrino fluxes can be  
sizable (see, for example, the discussion in Ref. [6]). Individual  
neutrinos emitted from the neutrino sphere can be described as  
coherent states. However, each emitted neutrino is related to every  
other emitted neutrino in an incoherent fashion. In other words,  
these different individual (or single) neutrino states have random  
relative phases, as is characteristic of a thermal emission process.  
The total neutrino field is properly a {\it mixed} ensemble of  
individual neutrino states. It is not a coherent many-body state.  
Accordingly, the total neutrino density matrix is an incoherent sum  
over each {\it single} neutrino density matrix.

For a single neutrino emitted at the neutrino sphere as a  
$\nu_\alpha$ (e.g., in flavor state $\alpha=e,\ \tau$ for the case of  
two-neutrino mixing) we can represent its state at some point above  
the neutrino sphere as
$$|\psi_{\nu_\alpha}\rangle=a_{1\alpha}(t)|\nu_1(t)\rangle+a_{2\alpha} 
(t)|\nu_2(t)\rangle,\eqno(9)$$
where $|\nu_1(t)\rangle$ and $|\nu_2(t)\rangle$ are the \  
instantaneous \ physical \ mass \ eigenstates \ of the full  
neutrino-propagation Hamiltonian, and $a_{1\alpha}(t)$ and  
$a_{2\alpha}(t)$ are the corresponding
complex amplitudes. Normalization requires that we take   
$|a_{1\alpha}(t)|^2+|a_{2\alpha}(t)|^2=1$. In these expressions the  
time, $t$, could be any evolutionary parameter (e.g., density,  
radius, etc.) along the neutrino's path from its creation position at  
the neutrino sphere to a point at radius $r$. The single neutrino  
density matrix is then given by
$$\eqalignno{|\psi_{\nu_\alpha}\rangle\langle\psi_{\nu_\alpha}|&=|a_{1 
\alpha}(t)|^2|\nu_1(t)\rangle\langle\nu_1(t)|+|a_{2\alpha}(t)|^2|\nu_2 
(t)\rangle\langle\nu_2(t)|\cr
&+a_{1\alpha}^*(t)a_{2\alpha}(t)|\nu_2(t)\rangle\langle\nu_1(t)|
+a_{1\alpha}(t)a_{2\alpha}^*(t)|\nu_1(t)\rangle\langle\nu_2(t)|.&(10)\cr}$$

The density matrix representing the mixed ensemble of single neutrino  
states all with momentum $\bf p$ can be written as the incoherent sum
$$\rho_{\bf p}d^3{\bf p}=\sum_{\alpha}  
dn_{\nu_\alpha}|\psi_{\nu_\alpha}\rangle\langle\psi_{\nu_\alpha}|.\eqno(11)$$
In this expression the sum runs over, for example, $\alpha=e,\ \tau$,  
while $dn_{\nu_\alpha}$ is the local differential number density of  
$\nu_\alpha$ neutrinos with momentum $\bf p$ in interval $d^3\bf p$.  
The local differential $\nu_\alpha$ neutrino number density at a  
point at radius $r$ above a neutrino sphere with radius $R_\nu$ is
$$dn_{\nu_\alpha}\approx n_{\nu_\alpha}^0  
f_{\nu_\alpha}(E_{\nu_\alpha})dE_{\nu_\alpha}\left({d\Omega_{\bf p}  
\over 4\pi}\right),\eqno(12a)$$
where $d\Omega_{\bf p}$ is the differential solid angle (pencil of  
directions) along the neutrino momentum $\bf p$ $(|{\bf p}|\approx  
E_{\nu_\alpha})$, $n_{\nu_\alpha}^0$ is the $\nu_\alpha$ neutrino  
number density at the neutrino sphere, and  
$f_{\nu_\alpha}(E_{\nu_\alpha})$ is the normalized $\nu_\alpha$  
energy distribution function. We can show [6] that a good  
approximation for $n_{\nu_\alpha}^0$ is 
$$n_{\nu_\alpha}^0\approx{L_{\nu_\alpha}\over\langle  
E_{\nu_\alpha}\rangle}{1\over \pi R_\nu^2c}, \eqno(12b)$$
where $L_{\nu_\alpha}$ is the luminosity in $\nu_\alpha$ neutrinos,  
$\langle E_{\nu_\alpha}\rangle$ is the average $\nu_\alpha$ neutrino  
energy, and $c$ is the speed of light. The normalized $\nu_\alpha$  
neutrino energy distribution function can be well approximated by
$$f_{\nu_\alpha}(E_{\nu_\alpha})\approx{1\over F_2(0)}{1\over  
T_{\nu_\alpha}^3} {E_{\nu_\alpha}^2\over  
\exp(E_{\nu_\alpha}/T_{\nu_\alpha})+1},\eqno(12c)$$
where the rank 2 Fermi integral with argument zero is $F_2(0)\approx  
1.803$, and where $T_{\nu_\alpha}$ is the $\nu_\alpha$ neutrino  
sphere temperature. The average $\nu_\alpha$ neutrino energy is  
related to the appropriate neutrino sphere temperature by
$\langle E_{\nu_\alpha}\rangle\approx 3.15T_{\nu_\alpha}$.

In the region of the supernova above the neutrino sphere, the range  
of the solid angle contribution allowed in Eq. (12a) is restricted to  
be within the solid angle subtended by the neutrino sphere as seen  
from a point at radius $r$. The geometrical arrangement of a neutrino  
sphere with radius $R_\nu$, a point above the neutrino sphere at  
radius $r$, and various neutrino paths are depicted in Fig. 2.

We can now write the full flavor-basis neutrino-propagation  
Hamiltonian as a sum of vacuum mass and electron background  
contributions, $H_e$, and neutrino background contributions,  
$H_{\nu\nu}$:
$$H=H_e+H_{\nu\nu},\eqno(13a)$$
where $H_{\nu\nu}$ represents the ensemble average over  
neutrino-neutrino interactions using the density matrix in Eq. (11).  
For a neutrino with energy $E_\nu$ and momentum $\bf p$ propagating  
radially outside the neutrino sphere we can write
$$H_{\nu\nu}=\sqrt{2}G_{\rm F}\int(1-\cos\theta_{\bf q})(\rho_{\bf  
q}-\bar\rho_{\bf q})d^3{\bf q},\eqno(13b)$$
where $\bar\rho_{\bf q}$ is the density matrix for antineutrinos with  
momentum $\bf q$ (defined in obvious analogy to $\rho_{\bf q}$ in Eq.  
[11]) and $\theta_{\bf q}$ is the angle between the direction of the  
propagating neutrino with momentum $\bf p$ and the directions of  
other neutrinos in the ensemble with momentum $\bf q$. We can  
generalize the expression for $H_{\nu\nu}$ in Eq. (13b) for  
non-radially propagating neutrinos by replacing $\cos\theta_{\bf q}$  
with ${\bf q\cdot p}/\{|{\bf q}||{\bf p}|\}$.

It is convenient to recast Eq. (13b) in the form
$$H_{\nu\nu}={1\over 2}\pmatrix{B&B_{e\tau}\cr B_{\tau e}&-B\cr}+  
{\sqrt{2}\over 2}G_{\rm F}\int(1-\cos\theta_{\bf q}){\rm  
Tr}(\rho_{\bf q}-\bar\rho_{\bf q})d^3{\bf q}.\eqno(14)$$
Note that the second term in this equation is simply proportional to  
the identity matrix, implying that it provides only an overall phase  
in the propagating neutrino state and can be ignored.

In the first term in Eq. (14) there are two contributions to the  
neutrino-propagation Hamiltonian, $B$ and $B_{e\tau}$ ($B_{\tau e}$),  
where
$$\eqalignno{B&=\sqrt{2}G_{\rm F}\int(1-\cos\theta_{\bf  
q})\{(\rho_{\bf q}-\bar\rho_{\bf q})_{ee}-(\rho_{\bf q}-\bar\rho_{\bf  
q})_{\tau\tau}\}d^3{\bf q},&(15a)\cr
B_{e\tau}&=2\sqrt{2}G_{\rm F}\int(1-\cos\theta_{\bf q})(\rho_{\bf  
q}-\bar\rho_{\bf q})_{e\tau}d^3{\bf q},&(15b)\cr
B_{\tau e}&=2\sqrt{2}G_{\rm F}\int(1-\cos\theta_{\bf q})(\rho_{\bf  
q}-\bar\rho_{\bf q})_{\tau e}d^3{\bf q},&(15c)\cr}$$
where, for example, by $(\rho_{\bf q})_{e\tau}$ we mean the matrix  
element of the density matrix operator, $\langle\nu_e|\rho_{\bf  
q}|\nu_\tau\rangle$, while by $(\bar\rho_{\bf q})_{e\tau}$ we mean  
$\langle\bar\nu_e|\rho_{\bf q}|\bar\nu_\tau\rangle$.

Here $B$ corresponds to the forward neutrino-neutrino  
exchange-scattering contributions to the neutrino effective mass.  
These contributions are the analogs of the $\nu_e$-$e$  
exchange-scattering term, $A$, in Eqs. (5) and (6). Generic Feynman  
graphs for these neutrino-neutrino exchange processes are shown in  
Fig. 3a for $\nu_e$-$\nu_e$ scattering and in Fig. 3b for  
$\nu_\tau$-$\nu_\tau$ scattering. We will later refer to $B$ as the  
``diagonal'' contribution of the neutrino background to the  
flavor-basis neutrino-propagation Hamiltonian.

The \ neutrino \ background \ also \ provides \ ``off-diagonal'' \  
terms in the flavor-basis neutrino-propagation Hamiltonian. These  
are, for example, the $B_{e\tau}$ and $B_{\tau e}$ terms above. They  
arise because the background neutrinos are not in flavor eigenstates  
[1]. We show graphically these contributions for $\nu_e$ and  
$\nu_\tau$ neutrinos with momenta $\bf p$ and $\bf q$ in Fig. 4. The  
corresponding diagonal and off-diagonal contributions to the  
flavor-basis antineutrino-propagation Hamiltonian from the neutrino  
background are $-B$ and $-B_{e\tau}\ (-B_{\tau e})$, respectively.

Considerable simplification in the evaluation of $B,\ B_{e\tau}\  
(B_{\tau e})$ can be realized by adroit attention to the phases in  
the expression for the single neutrino density matrix in Eq. (10).  
Note that the last two terms in Eq. (10) are cross terms. They have  
coefficients $a_1^*(t)a_2(t)$ and $a_1(t)a_2^*(t)$, respectively.  
Each cross term is proportional to a factor  
$\sim\exp[i\int\omega_{12}(t)dt]$, with $\omega_{12}$ the difference  
in the local neutrino flavor-oscillation frequencies of the two mass  
eigenstates $|\nu_1(t)\rangle$ and $|\nu_2(t)\rangle$. These  
oscillation frequencies are, in turn, dependent on the local density.

In both the early post-core-bounce shock re-heating epoch (time  
post-core-bounce $t_{\rm PB}\sim 0.1$--1 s) and in the hot  
bubble/$r$-process nucleosynthesis epoch ($t_{\rm PB}\sim 3$--15 s)  
the electron number density predominantly determines the neutrino  
flavor-oscillation frequency in the region just above the neutrino  
sphere [5, 6]. This is because the net neutrino number densities are  
negligible compared to the electron number densities ($n_e\sim  
10^{35}$ cm$^{-3}$) in this region [5, 6]. The Hamiltonian $H_e$ in  
Eq. (6) is by itself sufficient to determine the neutrino  
flavor-oscillation frequencies in this region. The local neutrino  
flavor-oscillation frequency difference in this case is  
$\omega_{12}\approx\Delta_{\rm eff}=  
\sqrt{(\Delta\cos2\theta-A)^2+\Delta^2\sin^22\theta}$.

Furthermore, for the cosmologically interesting range $\delta  
m^2=1$--100 eV$^2$, the electron number density near the neutrino  
sphere greatly exceeds the MSW resonance densities for neutrinos with  
energies $E_\nu\sim 10$ MeV. In this case the neutrino paths shown in  
Fig. 2 will always cross a region where the electron number densities  
dominate the neutrino number densities prior to entering the  
resonance region. Therefore the local neutrino flavor-oscillation  
frequency difference will be $\omega_{12}\approx\Delta_{\rm  
eff}\approx A=\sqrt{2}G_{\rm F}n_e$ along some part of every  
neutrino's path.

When taking the ensemble average over the neutrino background we necessarily
integrate $\rho_{\bf p}$ over neutrino momentum {\it directions} to a  
point at radius $r$. We thereby also average over the oscillating  
cross terms in Eq. (10). In addition, neutrinos with different  
momentum directions travel on paths with different lengths to arrive  
at a point at radius $r$. These different path lengths then give rise  
to different phases for the oscillating factor  
$\sim\exp[i\int\omega_{12}(t)dt]$ in the cross terms in Eq. (10). In  
fact it is clear from Fig. 2 that each neutrino path from the  
neutrino sphere to a point at radius $r$ will have a path length  
which depends on the polar angle. For neutrinos with momentum  
magnitude $|{\bf p}|$ each path with a different polar angle will  
have a different phase entering into the cross term coefficients of  
Eq. (10).

The phase difference $\delta\phi$ acquired in going through a region  
of electron number density $n_e$ with a difference in path length  
$\delta r$ is then
$$\delta\phi\approx\sqrt{2}G_{\rm F}n_e\delta r\approx  
642\left({n_e\over 10^{35}\ {\rm cm}^{-3}}\right)\left({\delta r\over  
1\ {\rm cm}}\right).\eqno(16)$$
Path length differences of order 1 cm give rise to phase differences  
of $\gg 2\pi$ ! It is obvious that the cross terms in the single  
neutrino density matrix in Eq. (10) vanish when averaged over all  
neutrino momentum directions. This argument can also be applied to  
the evaluation of $\bar\rho_{\bf p}$ since  
$\omega_{12}>\sqrt{2}G_{\rm F}n_e$ everywhere for  
antineutrinos. 

Clearly, we need only consider the first two terms of Eq. (10) in  
evaluating matrix elements of the density matrices $\rho_{\bf p}$ and  
$\bar\rho_{\bf p}$ (cf. Eq. [11]). This will allow considerable  
simplification in computation of $B$ and $B_{e\tau}\ (B_{\tau e})$  
from Eqs. (15a--c).

Failure to properly perform the angular part of the ensemble average  
would result in the retention of non-zero cross terms in the neutrino  
density matrix elements, Eqs. (15a--c). This would introduce a  
spurious, and unphysical, ``coherence'' in the treatment of the  
neutrino background. In fact, the angular part of the ensemble average
over the neutrino background is a key point in determining neutrino
flavor evolution in the region above the neutrino sphere in supernovae.

Note that $B_{e\tau}=B_{\tau e}$ since the terms in the ensemble  
averages,
Eqs. (15a--c), are all real and the Hamiltonian must be Hermitian.  
The terms in Eqs. (15a--c) are all real because of the vanishing of  
the cross terms in the momentum direction average over the single  
neutrino density matrices. The full flavor-basis Hamiltonian which  
includes both the electron and neutrino backgrounds is now
$$H=H_e+H_{\nu\nu}={1\over  
2}\pmatrix{-\Delta\cos2\theta+A+B&\Delta\sin2\theta+B_{e\tau}\cr
\Delta\sin2\theta+B_{e\tau}&\Delta\cos2\theta-A-B\cr}.\eqno(17a)$$
In analogy to the discussion preceding Eq. (6) we can rewrite this  
Hamiltonian as
$$H={\Delta_{\rm H}\over 2}\pmatrix{-\cos2\theta_{\rm  
H}&\sin2\theta_{\rm H}\cr\sin2\theta_{\rm H}&\cos2\theta_{\rm  
H}\cr}.\eqno(17b)$$
In this expression we have defined a full effective mixing angle,  
$\theta_{\rm H}$, which, in analogy to Eq. (1), gives the relations  
between the flavor basis and the instantaneous mass basis including  
the effects of both the electron and neutrino backgrounds:
$$\eqalignno{|\nu_e(t)\rangle&=\cos\theta_{\rm  
H}(t)|\nu_1(t)\rangle+\sin\theta_{\rm H}(t)|\nu_2(t)\rangle,&(18a)\cr
|\nu_\tau(t)\rangle&=-\sin\theta_{\rm  
H}(t)|\nu_1(t)\rangle+\cos\theta_{\rm  
H}(t)|\nu_2(t)\rangle.&(18b)\cr}$$
We have defined $\Delta_{\rm H}$ as
$$\Delta_{\rm H}\equiv  
\sqrt{(\Delta\cos2\theta-A-B)^2+(\Delta\sin2\theta+B_{e\tau})^2}.\eqno 
(19)$$ 

The full effective mixing angle satisfies
$$\sin2\theta_{\rm H}={\Delta\sin2\theta_{\rm  
H}+B_{e\tau}\over\sqrt{(\Delta\cos2\theta-A-B)^2+(\Delta\sin2\theta+B_ 
{e\tau})^2}},\eqno(20a)$$
$$\cos2\theta_{\rm H}={\Delta\cos2\theta_{\rm  
H}-A-B\over\sqrt{(\Delta\cos2\theta-A-B)^2+(\Delta\sin2\theta+B_{e\tau 
})^2}}.\eqno(20b)$$
Note that in the absence of a neutrino background $\Delta_{\rm  
H}=\Delta_{\rm eff}$ and $\theta_{\rm H}=\theta_n$. The corresponding  
expressions for the full effective mixing angle, $\bar\theta_{\rm  
H}$, in the antineutrino sector are obtained by replacing $A,\ B$,  
and $B_{e\tau}$ with $-A,\ -B$, and $-B_{e\tau}$, respectively.

Since the cross terms in the single neutrino density matrix will give  
no contribution to the ensemble average, we can write a reduced  
expression for the single neutrino density matrix in terms of 
flavor-basis eigenbras and eigenkets:
$$\eqalignno{(|\psi_{\nu_\alpha}\rangle\langle\psi_{\nu_\alpha}|)_{\rm
reduced}&=\{{1\over 2}-[{1\over 2}-|a_{1\alpha}(t)|^2]\cos2\theta_{\rm  
H}(t)\}|\nu_e\rangle\langle\nu_e|\cr
&+\{{1\over 2}-[{1\over 2}+|a_{1\alpha}(t)|^2]\cos2\theta_{\rm  
H}(t)\}|\nu_\tau\rangle\langle\nu_\tau|\cr
&+[{1\over 2}-|a_{1\alpha}(t)|^2]\sin2\theta_{\rm  
H}(t)(|\nu_e\rangle\langle\nu_\tau|+|\nu_\tau\rangle\langle\nu_e|).&(2 
1)\cr}$$

With this form for the single neutrino density matrix, it is  
straightforward to evaluate flavor-basis matrix elements of the  
density matrix operator. For example, the expressions in Eqs.  
(15a--c) become
$$\eqalignno{B=-\sqrt{2}G_{\rm F}\sum_\alpha\int(1-\cos\theta_{\bf  
q})&\{[1-2|a_{1\alpha}(t)|^2]\cos2\theta_{\rm H}(t)dn_{\nu_\alpha}\cr
&-[1-2|\bar a_{1\alpha}(t)|^2]\cos2\bar\theta_{\rm  
H}(t)dn_{\bar\nu_\alpha}\},&(22a)\cr}$$
$$\eqalignno{B_{e\tau}=\sqrt{2}G_{\rm  
F}\sum_\alpha\int(1-\cos\theta_{\bf  
q})&\{[1-2|a_{1\alpha}(t)|^2]\sin2\theta_{\rm H}(t)dn_{\nu_\alpha}\cr
&-[1-2|\bar a_{1\alpha}(t)|^2]\sin2\bar\theta_{\rm  
H}(t)dn_{\bar\nu_\alpha}\}.&(22b)}$$
In these expressions $a_{1\alpha}(t)$ is the amplitude to be in the  
instantaneous mass eigenstate $|\nu_1(t)\rangle$ for an individual  
neutrino of momentum $\bf q$ which was created at the neutrino sphere  
($t=0$) in flavor eigenstate $|\nu_\alpha\rangle$. Likewise, $\bar  
a_{1\alpha}(t)$ is the amplitude to be in the instantaneous mass  
eigenstate $|\bar\nu_1(t)\rangle$ for an antineutrino of momentum  
$\bf q$ created at the neutrino sphere in flavor eigenstate  
$|\bar\nu_\alpha\rangle$. The expressions $dn_{\nu_\alpha}$ and  
$dn_{\bar\nu_\alpha}$ are as given in Eq. (12a), e.g.,  
$dn_{\nu_\alpha}\approx  
n_{\nu_\alpha}^0f_{\nu_\alpha}(E_{\nu_\alpha})dE_{\nu_\alpha}(d\Omega_ 
{\bf q}/ 4\pi)$.

It remains to evaluate these expressions for the particular  
conditions (electron density run and neutrino distribution functions)  
which obtain for the shock re-heating and hot bubble/$r$-process  
nucleosynthesis epochs.
\vskip .2in
\centerline{\bf III. Neutrino Flavor Transformation in the Supernova  
Environment}
\vskip .2in
In this section we examine neutrino flavor transformation in the  
region above the neutrino sphere in models of post-core-bounce Type  
II supernovae. There are several aspects of the problem of neutrino  
flavor transformation in supernovae which are significantly different  
from conventional computations of MSW flavor conversion in the sun.  
Foremost among these is the necessity of treating the neutrino  
background. In addition, the geometry of neutrino emission from a  
neutrino sphere in a supernova is quite different from the solar  
case, where the neutrino source is distributed throughout the core.

Bearing these points in mind, we can formally transform the full  
flavor-basis Hamiltonian in Eqs. (17a \& b) to the basis of the  
instantaneous mass eigenstates $|\nu_1(t)\rangle$ and  
$|\nu_2(t)\rangle$. The Schr\"odinger equation for the time evolution  
of the amplitudes $a_{1\alpha}(t)$ and $a_{2\alpha}(t)$ (see Eq. [9])  
in this basis is then,
$$i\pmatrix{\dot a_{1\alpha}(t)\cr\dot  
a_{2\alpha}(t)\cr}=\pmatrix{-\Delta_{\rm H}(t)/2&-i\dot\theta_{\rm  
H}(t)\cr i\dot\theta_{\rm H}(t)&\Delta_{\rm  
H}(t)/2\cr}\pmatrix{a_{1\alpha}(t)\cr a_{2\alpha}(t)\cr},\eqno(23)$$
where $\dot a_{1\alpha}(t)=da_{1\alpha}(t)/dt$, $\dot  
a_{2\alpha}(t)=da_{2\alpha}(t)/dt$, and $\dot\theta_{\rm  
H}(t)=d\theta_{\rm H}(t)/dt$. In this expression we follow the  
treatment of neutrino propagation and flavor transformation in Ref.  
[8]. Eq. (23) represents a set of {\it nonlinear} first order  
differential equations for the amplitudes $a_{1\alpha}(t)$ and  
$a_{2\alpha}(t)$. The nonlinearity arises since, in general,  
$\Delta_{\rm H}$ and the full effective mixing angle $\theta_{\rm H}$  
each depend on the neutrino background contributions $B$ and  
$B_{e\tau}$ (Eqs. [19], [20a \& b]). In turn, $B$ and $B_{e\tau}$  
depend on the amplitudes $a_{1\alpha}(t)$ as in Eqs. (22a \& b).

The time evolution of the full effective mixing angle can be found  
from Eqs. (20a \& b) to be
$$\dot\theta_{\rm H}(t)={\dot  
B_{e\tau}(\Delta\cos2\theta-A-B)+(\Delta\sin2\theta+B_{e\tau})(\dot  
A+\dot B)\over  
2[(\Delta\cos2\theta-A-B)^2+(\Delta\sin2\theta+B_{e\tau})^2]},\eqno(24 
)$$
where $\dot A=dA/dt$, $\dot B=dB/dt$, and $\dot  
B_{e\tau}=dB_{e\tau}/dt$.

We can define an ``adiabaticity parameter'' $\gamma(t)$:
$$\gamma(t)\equiv {\Delta_{\rm H}(t)\over 2|\dot\theta_{\rm  
H}(t)|}.\eqno(25)$$
Clearly, the neutrino mass eigenstate evolution is well approximated  
as being adiabatic when $\gamma(t)\gg 1$. Of course, if  
$\dot\theta_{\rm H}=0$, the neutrino mass eigenstate evolution is  
completely adiabatic, as can be seen directly from Eq. (23).

The adiabaticity parameter generally satisfies $\gamma(t)\gg 1$ well  
away from resonance regions (neutrino mass-level-crossing regions).  
However, neutrino flavor conversion probabilities depend crucially on  
$\gamma(t)$ at resonance. We shall denote the value of the  
adiabaticity parameter at resonance as $\gamma(t_{\rm res})$.  
Resonance occurs when
$$\Delta\cos2\theta=A+B.\eqno(26)$$
We denote the position of this level-crossing point, or resonance, by  
$t_{\rm res}$. At resonance,
$$\gamma(t_{\rm res})={(\Delta\sin2\theta+B_{e\tau})^2\over|\dot  
A+\dot  
B|}={(\Delta\sin2\theta+B_{e\tau})^2\over\Delta\cos2\theta}|{d\ln(A+B) 
\over dt}|^{-1}_{t_{\rm res}}.\eqno(27)$$

The Landau-Zener probability for the neutrino to jump from one mass  
eigenstate to the other in the course of transversing a resonance  
region is [8],
$$P_{\rm LZ}\approx\exp[-{\pi\over 2}\gamma(t_{\rm res})].\eqno(28)$$
Unlike the case for solar neutrinos, this expression is always  
sufficient for calculating neutrino flavor transformation in  
supernovae [5, 6]. The Landau-Zener formula Eq. (28) is inapplicable  
for solar neutrino flavor
conversion when, for example, neutrinos are created close to their  
resonance positions. This never occurs in supernovae, where neutrinos  
originate on the neutrino sphere. The neutrino sphere is always well
away from the resonance region for the cases we will consider. In addition, solar  
neutrinos can experience double level crossings when they are created  
at densities below their resonance density. This does not occur in  
the post-core-bounce supernova environment. 

The very small vacuum mixing angles we shall consider for neutrino  
flavor conversion in supernovae imply narrow resonance regions.  
Narrow resonance regions, together with the generally large density  
scale heights (0.5--50 km) characterisitic of the region above the  
neutrino sphere [5, 6], imply that the first order Landau-Zener jump  
probability expression in Eq. (28) is always adequate [5, 6]. By  
first order jump probability we mean that we approximate the density  
profile as linear across the resonance region.

It is obvious in Eqs. (24--27) that we recover the pure  
electron-driven neutrino flavor conversion case when the neutrino  
background disappears (i.e., $B$ and $B_{e\tau}$ vanish everywhere).  
The neutrino background influences neutrino flavor evolution through  
resonances in two ways.

First, the diagonal contribution of the neutrino background, $B$,  
essentially shifts the position of the resonance from the case where  
only the electron contribution, $A$, is present. This is evident from  
Eq. (26). The diagonal contribution of the neutrino background also  
alters the density scale height of weak interaction scattering  
targets at resonance. The density scale height of weakly interacting  
targets ($|d\ln n/dr|^{-1}$ following Eq. [7] in Ref. [6]) is the  
$|d\ln(A+B)/dt|^{-1}$ term in Eq. (27).

The off-diagonal contribution of the neutrino background,  
$B_{e\tau}$, has the effect of altering the adiabaticity of the  
neutrino flavor evolution at resonance. This is clear from Eq. (27),  
where $B_{e\tau}$ appears in the expression for $\gamma(t_{\rm  
res})$. If $\Delta\sin2\theta\gg|B_{e\tau}|$ then the off-diagonal  
neutrino background contribution will have little influence on the  
adiabaticity of neutrino flavor evolution.

However, the diagonal and off-diagonal contributions of the neutrino  
background influence neutrino flavor evolution in a nonlinear manner,  
as outlined above. Not only are $B$ and $B_{e\tau}$ determined by the  
local neutrino distribution functions, but the local neutrino  
distribution functions are also dependent, in general, on the  
detailed history of neutrino flavor transformation.

The crux of the problem of treating the nonlinear effects of the  
neutrino background is the computation of $B$ and $B_{e\tau}$ for the  
particular local neutrino distribution functions which obtain in the  
supernova environment. This will be evident if we discuss a simple  
iterative procedure for computing neutrino flavor transformation at  
resonance in the presence of a neutrino background.

We can employ the Landau-Zener transformation probability in Eq. (28)  
to estimate the neutrino flavor conversion probability for a neutrino  
propagating through a resonance with the following simple procedure.
We choose a vacuum mass-squared difference $\delta m^2$ and a vacuum  
mixing angle $\theta$ (equivalently, $\sin^22\theta$) for a  
propagating neutrino of energy $E_{\rm R}$.

(1.) To begin with, we assume that $B_{e\tau}=0$. We use $\delta m^2$  
and $\sin^22\theta$, along with $B_{e\tau}=0$, in Eqs. (20a \& b) to  
get a zero-order estimate for $\cos2\theta_{\rm H}$,  
$\sin2\theta_{\rm H}$, $\cos2\bar\theta_{\rm H}$, and  
$\sin2\bar\theta_{\rm H}$. Note that the value of $A$ and $B$ which  
enter into the expressions for $\cos2\theta_{\rm H}$,  
$\sin2\theta_{\rm H}$, $\cos2\bar\theta_{\rm H}$, and  
$\sin2\bar\theta_{\rm H}$ are their values at the resonance position,  
$A(t_{\rm res})$ and $B(t_{\rm res})$. In this case we can replace  
$A+B$ by $(\delta m^2/2E_{\rm R})\cos2\theta$ wherever it occurs. Eqs.  
(20a \& b) with $B_{e\tau}=0$ can then be written as:
$$\eqalignno{\sin2\theta_{\rm H}&={\tan2\theta\over\sqrt{(1-E/E_{\rm  
R})^2+\tan^22\theta}},&(29a)\cr
\cos2\theta_{\rm H}&={1-E/E_{\rm R}\over\sqrt{(1-E/E_{\rm  
R})^2+\tan^22\theta}},&(29b)\cr
\sin2\bar\theta_{\rm H}&={\tan2\theta\over\sqrt{(1+E/E_{\rm  
R})^2+\tan^22\theta}},&(29c)\cr
\cos2\bar\theta_{\rm H}&={1+E/E_{\rm R}\over\sqrt{(1+E/E_{\rm  
R})^2+\tan^22\theta}}.&(29d)\cr}$$

(2.) We employ these approximations for the full effective mixing  
angle to obtain estimates for $B$ in Eq. (22a).

(3.) So far we have not specified the resonance position. We now use  
$A$ and the estimate of $B$ from Step (2) to estimate the resonance  
position through $(\delta m^2/2E_{\rm R})\cos2\theta=A+B$. Note that  
$A$ and $B$ are position dependent.

(4.) With the resonance position from Step (3) we use Eq. (22b) to  
estimate $B_{e\tau}$.

(5.) With this estimate for $B_{e\tau}$ we now can re-estimate the  
full effective mixing angle using
$$\eqalignno{
\sin2\theta_{\rm H}&={(\delta m^2/2E_\nu)\sin2\theta+B_{e\tau}
\over\sqrt{[(\delta m^2/2E_\nu)-(\delta m^2/2E_{\rm R})]^2
\cos^22\theta+[(\delta  
m^2/2E_\nu)\sin2\theta+B_{e\tau}]^2}},&(30a)\cr
\cos2\theta_{\rm H}&={[(\delta m^2/2E_\nu)-(\delta m^2/2E_{\rm  
R})]\cos2\theta\over\sqrt{[(\delta m^2/2E_\nu)-(\delta m^2/2E_{\rm  
R})]^2\cos^22\theta+[(\delta  
m^2/2E_\nu)\sin2\theta+B_{e\tau}]^2}},&(30b)\cr
\sin2\bar\theta_{\rm H}&={(\delta m^2/  
2E_\nu)\sin2\theta-B_{e\tau}\over\sqrt{[(\delta m^2/2E_\nu)+(\delta  
m^2/2E_{\rm R})]^2\cos^22\theta+[(\delta  
m^2/2E_\nu)\sin2\theta-B_{e\tau}]^2}},&(30c)\cr
\cos2\bar\theta_{\rm H}&={[(\delta m^2/2E_\nu)+(\delta m^2/2E_{\rm  
R})]\cos2\theta\over\sqrt{[(\delta m^2/2E_\nu)+(\delta m^2/2E_{\rm  
R})]^2\cos^22\theta+[(\delta  
m^2/2E_\nu)\sin2\theta-B_{e\tau}]^2}}.&(30d)\cr}$$

(6.) We iterate by returning to Step (2) and re-evaluating $B$.

This procedure must be continued until $B,\ B_{e\tau},\ \theta_{\rm  
H}$ and the resonance position $(t_{\rm res})$ converge. Because of  
the dependence of $B$ and $B_{e\tau}$ on the flavor evolution  
histories of all neutrinos in the ensemble, convergence of this  
procedure is, in general, problematic. However, if neutrino flavor  
evolution is adiabatic then the complication of prior histories is  
eliminated and the above procedure converges rapidly for the  
conditions which obtain in the region above the neutrino sphere in  
Type II supernovae. For nonadiabatic neutrino flavor evolution the
above procedure, though more laborious, still gives good estimates of
the effects of the  
neutrino background. We shall begin by discussing the case of  
adiabatic neutrino flavor evolution.
\vskip .2in
\centerline{\bf IIIa.) Adiabatic Neutrino Flavor Evolution}
\vskip .2in
Consider the flavor evolution of antineutrinos. It is generally true  
everywhere above the neutrino sphere that the contributions of the  
electrons and neutrinos satisfy $A+B>0$. This is true because $n_e$  
is everywhere greater than the net neutrino number densities for any  
neutrino flavor [5,6]. For an antineutrino emitted from the neutrino  
sphere in the $|\bar\nu_e\rangle$ flavor eigenstate, it is evident that
$|\bar a_{1e}(t)|^2\approx 1$ and $|\bar a_{1\tau}(t)|^2\approx 0$ for
all $t$. The effective mass-squared difference for two antineutrino
mass eigenstates always {\it increases} with density and there is no
mass level crossing. The adiabatic approximation for the evolution of
the antineutrino  mass eigenstates is always good.

The situation is more complicated for neutrinos. However, the  
approximation of adiabatic evolution of the neutrino mass eigenstates  
is a particularly simple case to treat in the supernova. A neutrino  
created in a flavor eigenstate $|\nu_\alpha\rangle$ at the neutrino  
sphere is very nearly in a mass eigenstate because of the large  
electron number density there. Subsequent adiabatic evolution then  
implies that, for example, $|a_{1e}(t)|^2=0$ and $|a_{1\tau}(t)|^2=1$  
for all $t$ (likewise, $|a_{2e}(t)|^2=1$ and $|a_{2\tau}(t)|^2=0$ for  
all $t$). In this case the expressions for the neutrino background  
contributions, Eqs. (22a \& b), become
$$B\approx -\sqrt{2}G_{\rm F}\int(1-\cos\theta_{\bf  
q})[\cos2\theta_{\rm  
H}(t)(dn_{\nu_e}-dn_{\nu_\tau})+\cos2\bar\theta_{\rm  
H}(t)(dn_{\bar\nu_e}-dn_{\bar\nu_\tau})],\eqno(31a)$$
$$B_{e\tau}\approx \sqrt{2}G_{\rm F}\int(1-\cos\theta_{\bf  
q})[\sin2\theta_{\rm  
H}(t)(dn_{\nu_e}-dn_{\nu_\tau})+\sin2\bar\theta_{\rm  
H}(t)(dn_{\bar\nu_e}-dn_{\bar\nu_\tau})].\eqno(31b)$$

The evaluation of Eqs. (31a \& b) for particular neutrino  
distribution functions is straightforward so long as the adiabatic  
approximation obtains. To begin with, consider the computation of $B$  
from Eq. (31a) in the limit where $B_{e\tau}=0$. The result so  
obtained will be valid if we can later show that $|B_{e\tau}|\ll  
(\delta m^2/2E_{\rm R})\sin2\theta$.

With the approximation that $B_{e\tau}$ is small the integrals over  
the neutrino distribution functions $dn_{\nu_e},\ dn_{\nu_\tau},\  
dn_{\bar\nu_e}$, and $dn_{\bar\nu_\tau}$ can be separated into an  
angular part and an energy part. This is due to the fact that when  
$B_{e\tau}$ is small $\theta_{\rm H}$ and $\bar\theta_{\rm H}$  
essentially become functions of energy alone. For a radially  
propagating neutrino, the angular part of the integral in Eq. (31a)  
then becomes,
$$\int(1-\cos\theta_{\bf q}){d\Omega_{\bf q}\over 4\pi}={1\over  
2}\int_0^{\theta_0}(1-\cos\theta)\sin\theta d\theta={1\over  
4}[1-\sqrt{1-(R_\nu/r)^2}\,]^2.\eqno(32a)$$
In this equation $r$ is the radius of the point at which we evaluate  
$B$ and $\theta_0$ is the polar angle of the limb of the neutrino  
sphere as seen from this point. Frequently we are interested in  
regions sufficiently distant from the neutrino sphere that we can take
$r\gg R_\nu$. In this limit, the radial neutrino path to the point at
radius $r$ is a good representation of all neutrino paths to that
point, and we can  
approximate
$$\int(1-\cos\theta_{\bf q}){d\Omega_{\bf q}\over 4\pi}\approx  
{1\over 16}{R_\nu^4\over r^4}.\eqno(32b)$$
It is obvious from this expression that the diagonal contribution of  
the neutrino background is sensitive to position.

The integration of the remaining energy dependent terms in Eq. (31a)  
is simple if we employ the approximate energy spectra in Eq. (12c).  
The energy part of Eq. (31a) is then,
$$\eqalignno{\int\cos2\theta_{\rm H}f_{\nu_\alpha}(E_{\nu_\alpha})  
dE_{\nu_\alpha}&\approx F_{\nu}(\theta,E_{\rm  
R}/T_{\nu_\alpha}),&(33a)\cr
\int\cos2\bar\theta_{\rm H}f_{\bar\nu_\alpha}(E_{\bar\nu_\alpha})  
dE_{\bar\nu_\alpha}&\approx F_{\bar\nu}(\theta,E_{\rm  
R}/T_{\bar\nu_\alpha}),&(33b)\cr}$$
where we define the neutrino spectral integrals as,
$$\eqalignno{F_\nu(\theta,x_{\rm R})&\equiv{1\over  
F_2(0)}\int_0^\infty{1-x/x_{\rm R}\over\sqrt{(1-x/x_{\rm  
R})^2+\tan^22\theta}}{x^2\over \exp(x)+1}dx,&(34a)\cr
F_{\bar\nu}(\theta,x_{\rm R})&\equiv{1\over  
F_2(0)}\int_0^\infty{1+x/x_{\rm R}\over\sqrt{(1+x/x_{\rm  
R})^2+\tan^22\theta}}{x^2\over \exp(x)+1}dx.&(34b)\cr}$$
Clearly, for $\tan2\theta\ll 1$, $F_{\bar\nu}(\theta,x_{\rm  
R})\approx 1$.
Here $E_{\rm R}$ is the energy corresponding to a neutrino at  
resonance at radius $r$.

With these definitions, and for small $B_{e\tau}$, we can reduce Eq.  
(31a) for $B$ to,
$$\eqalignno{B\approx-\sqrt{2}G_{\rm  
F}{[1-\sqrt{1-R_\nu^2/r^2}]^2\over 4}
&[n_{\nu_e}^0F_\nu(\theta,E_{\rm  
R}/T_{\nu_e})-n_{\nu_\tau}^0F_\nu(\theta, E_{\rm R}/T_{\nu_\tau})\cr
&+n_{\bar\nu_e}^0F_{\bar\nu}(\theta,E_{\rm  
R}/T_{\bar\nu_e})-n_{\bar\nu_\tau}^0F_{\bar\nu}(\theta,E_{\rm  
R}/T_{\bar\nu_\tau})],&(35)\cr}$$
where $n_{\nu_e}^0,\ n_{\nu_\tau}^0,\ n_{\bar\nu_e}^0$, and  
$n_{\bar\nu_\tau}^0$ are the appropriate neutrino or antineutrino  
number densities at the neutrino sphere as in Eq. (12b). This  
zero-order expression for $B$ is to be used in Step (2) in the  
iterative procedure outlined above. To proceed further requires that  
we estimate $B_{e\tau}$.

The angular integration for Eq. (31b) is the same as for Eq. (31a).  
In performing the angular integration in Eq. (31b) we will again  
assume that $B_{e\tau}$ is small. The energy dependent integrals in  
Eq. (31b) can be written as:
$$\eqalignno{\int\sin2\theta_{\rm H}f_{\nu_\alpha}(E_{\nu_\alpha})  
dE_{\nu_\alpha}&\approx G_{\nu}(\theta,E_{\rm  
R}/T_{\nu_\alpha}),&(36a)\cr
\int\sin2\bar\theta_{\rm H}f_{\bar\nu_\alpha}(E_{\bar\nu_\alpha})  
dE_{\bar\nu_\alpha}&\approx G_{\bar\nu}(\theta,E_{\rm  
R}/T_{\bar\nu_\alpha}).&(36b)\cr}$$
In like manner to Eqs. (34a \& b) we define,
$$\eqalignno{G_\nu(\theta,x_{\rm R})&\equiv{1\over  
F_2(0)}\int_0^\infty{\tan2\theta\over\sqrt{(1-x/x_{\rm  
R})^2+\tan^22\theta}}{x^2\over \exp(x)+1}dx,&(37a)\cr
G_{\bar\nu}(\theta,x_{\rm R})&\equiv{1\over  
F_2(0)}\int_0^\infty{\tan2\theta\over\sqrt{(1+x/x_{\rm  
R})^2+\tan^22\theta}}{x^2\over \exp(x)+1}dx,&(37b)\cr}$$
where the notation is as in Eqs. (34a \& b).

Finally, we can utilize Eqs. (36a--37b) to give an approximate  
expression for $B_{e\tau}$,
$$\eqalignno{B_{e\tau}\approx\sqrt{2}G_{\rm  
F}{[1-\sqrt{1-R_\nu^2/r^2}]^2\over 4}
&[n_{\nu_e}^0G_\nu(\theta,E_{\rm  
R}/T_{\nu_e})-n_{\nu_\tau}^0G_\nu(\theta, E_{\rm R}/T_{\nu_\tau})\cr
&+n_{\bar\nu_e}^0G_{\bar\nu}(\theta,E_{\rm  
R}/T_{\bar\nu_e})-n_{\bar\nu_\tau}^0G_{\bar\nu}(\theta,E_{\rm  
R}/T_{\bar\nu_\tau})].&(38)\cr}$$
The notation in this equation is the same as in Eq. (35). The  
approximations for $B$ and $B_{e\tau}$ in Eqs. (35) and (38),  
respectively, are valid when,
$|B_{e\tau}|/(\delta m^2/2E_{\rm R})\sin2\theta\ll 1$.

Note that the integrand in the expression for $G_{\nu}(\theta,x_{\rm  
R})$ in Eq. (37a) contains a factor, $\sin\theta_{\rm H}\approx  
\tan2\theta/\sqrt{(1-x/x_{\rm  
R})^2+\tan^22\theta}=\tan2\theta/\sqrt{(1-E_\nu/E_{\rm  
R})^2+\tan^22\theta}$, which is sharply peaked at $E_{\nu}=E_{\rm R}$  
for small vacuum mixing angles. In Fig. 5 we plot $\sin2\theta_{\rm  
H}$ as a function of $E_\nu/E_{\rm R}$ for three values of the vacuum  
mixing angle. The dotted line in this figure corresponds to  
$\tan2\theta=10^{-3}$. The dashed line corresponds to  
$\tan2\theta=10^{-2}$, while the solid line is for $\tan2\theta=0.1$.  
Since the factor $\sin2\theta_{\rm H}$ appears in the integration  
over the neutrino energy spectrum we can see easily that the smaller  
the vacuum mixing angle, the smaller will be the fraction of the  
total number density of neutrinos which contribute to $B_{e\tau}$.  
The physical interpretation of this is clear: the neutrinos which  
make the largest contribution to the off-diagonal neutrino background
terms are those which have the largest full effective mixing angles at
the position under consideration at radius $r$. These are the neutrinos
which have energies close to $E_{\rm R}$.

With the iterative procedure outlined above we can estimate $B$ and  
$B_{e\tau}$ for adiabatic neutrino flavor evolution in both the shock  
re-heating epoch and the hot bubble/$r$-process nucleosynthesis  
epoch. Refs. [5] and [6] give detailed expositions of the expected  
neutrino emission parameters for these epochs. Typical neutrino  
luminosities for the shock re-heating epoch at $t_{\rm PB}\approx  
0.15$ s (see the discussion in Ref. [5]) are $L_{\nu_e}\approx  
L_{\bar\nu_e}\approx L_{\nu_{\tau(\mu)}}\approx  
L_{\bar\nu_{\tau(\mu)}}\approx 5\times 10^{52}$ erg s$^{-1}$. The  
neutrino sphere radius at this epoch is $R_{\nu}\approx 50$ km, while  
the average neutrino energies are $\langle E_{\nu_e}\rangle\approx 9$  
MeV, $\langle E_{\bar\nu_e}\rangle\approx 12$ MeV, and $\langle  
E_{\nu_{\tau(\mu)}}\rangle\approx\langle  
E_{\bar\nu_{\tau(\mu)}}\rangle\approx 20$ MeV. By contrast, in the  
later hot bubble/$r$-process nucleosynthesis epoch ($t_{\rm  
PB}\approx 5$ s) the neutrino liminosities are $L_{\nu_e}\approx
L_{\bar\nu_e}\approx L_{\nu_{\tau(\mu)}}\approx  
L_{\bar\nu_{\tau(\mu)}}\approx 10^{51}$ erg s$^{-1}$, while the
neutrino sphere is at radius $R_\nu\approx 10$ km. The average neutrino
energies for this epoch are $\langle E_{\nu_e}\rangle\approx 11$  
MeV, $\langle E_{\bar\nu_e}\rangle\approx 16$ MeV, and $\langle  
E_{\nu_{\tau(\mu)}}\rangle\approx\langle  
E_{\bar\nu_{\tau(\mu)}}\rangle\approx 25$ MeV.

As Ref. [5] shows, for a substantial enhancement in shock re-heating  
$\nu_\tau$ (or $\nu_\mu$) neutrinos with energies $E_\nu\approx 35$  
MeV must be efficiently transformed into $\nu_e$ neutrinos in the  
region behind the stalled shock. Ref. [6] shows that neutrinos with  
energies $E_\nu\approx 25$ MeV are the most important in determining  
the electron fraction, $Y_e$, in the hot bubble/$r$-process  
nucleosynthesis epoch.

For shock re-heating enhancement we {\it must} have adiabatic  
transformation of neutrinos with energies $E_\nu\approx 35$ MeV. In  
the hot bubble/$r$-process nucleosynthesis epoch adiabatic  
transformation is not {\it necessary} to drive the material too  
proton rich for $r$-process nucleosynthesis to occur $(Y_e>0.5)$. In  
fact Ref. [6] shows that $\nu_{\tau(\mu)}\rightleftharpoons\nu_e$  
flavor conversion efficiencies as small as $\sim 30$\% for neutrinos  
with energies $E_\nu\approx 25$ MeV will suffice to drive $Y_e\ge  
0.5$. Nevertheless, for large enough vacuum mixing angles, adiabatic  
transformation of neutrinos with $E_\nu\approx 25$ MeV will occur in  
some regions of the $(\delta m^2,\ \sin^22\theta)$ plot (Fig. 2 in  
Ref. [6]).

Consider adiabatic neutrino flavor conversion specifically for  
$E_\nu=35$ MeV in the shock re-heating epoch and $E_\nu=25$ MeV in  
the hot bubble/$r$-process nucleosynthesis epoch. For comparison, we  
first present values of $\delta m^2$ and $\sin^22\theta$ which give  
an adiabaticity parameter $\gamma=3$ for the bare electron number  
density distributions relevant for these epochs. In Figs. 6 and 7,  
the solid contour lines for $\gamma=3$ correspond to these values of  
$\delta m^2$ and $\sin^22\theta$ for the representative conditions in  
the shock re-heating and hot bubble/$r$-process nucleosynthesis  
epochs, respectively.

It should be noted that the adiabatic approximation will be valid  
over the whole range of neutrino energies implicit in the neutrino  
distribution functions entering into the expressions for $B$ and  
$B_{e\tau}$. Neutrinos with energies $E_\nu<E_{\rm R}$ will propagate  
through resonances {\it prior} to reaching the resonant position for  
the specific example neutrino energy under discussion (either $E_{\rm  
R}=35$ MeV or $E_{\rm R}=25$ MeV). It is a general feature of the  
density scale height above the neutrino sphere that the neutrinos  
with energies $E_\nu<E_{\rm R}$ will experience adiabatic flavor  
evolution through their resonances as long as neutrinos with energy  
$E_{\rm R}$ go through the resonance adiabatically [5, 6]. Background  
neutrinos with energies $E_\nu>E_{\rm R}$ will not have gone through  
resonances and therefore evolve adiabatically {\it prior} to arriving  
at the resonance position for a neutrino with energy $E_{\rm R}$. We  
conclude that the expressions for $B$ and $B_{e\tau}$ in Eqs. (31a \&
b) are appropriate for the example under consideration.

Using the iterative procedure outlined above we can calculate the  
true adiabatic parameter, $\gamma(t_{\rm res})$, including the  
neutrino background contributions. We show the new contour lines for  
$\gamma=3$ as dotted lines in Figs. 6 and 7 for the respective epochs.  
We can easily see that the neutrino background has a completely  
negligible effect on adiabaticity at resonance along the solid  
$\gamma=3$ contour line in Fig. 6. The new contour line for  
$\gamma=3$ in Fig. 6 is indistinguishable from the one calculated for  
the bare electron number density. The new contour line for $\gamma=3$  
in Fig. 7 moves a little bit to the right of the solid line, but the  
neutrino background effects are also evidently small.

Any neutrino mixing parameters $\delta m^2$ and $\sin^22\theta$ which  
are to the right of the $\gamma=3$ contour lines in Figs. 6 and 7  
correspond to larger values of $\gamma$ for the specific example  
neutrino energies under discussion. For a given $\delta m^2$ the  
ratio $|B_{e\tau}|/(\delta m^2/ 2E_{\rm R})\sin2\theta$ will {\it  
decrease} as $\sin^22\theta$ and, hence, $\gamma$ {\it increases}.  
The off-diagonal neutrino background contribution will have a  
negligible effect on neutrino flavor conversion everywhere to the  
right of the contour lines in Figs. 6 and 7. Likewise, $B$ is roughly  
constant for a given $\delta m^2$ as $\sin^22\theta$ and, hence,  
$\gamma$ is increased. The diagonal contribution of the neutrino  
background produces a negligible alteration in the computed flavor  
conversion efficiencies everywhere to the right of the contour lines  
in Figs. 6 and 7.

We have also examined adiabatic neutrino flavor conversion in  
supernovae for a range of neutrino energies. We can conclude that the  
neutrino background, specifically $B$ and $B_{e\tau}$, will not  
result in any modification of the results of Refs. [5] and [6]  
whenever {\it adiabatic} neutrino flavor evolution is at issue.
\vskip .2in
\centerline{\bf IIIb.) Nonadiabatic Neutrino Flavor Evolution}
\vskip .2in
The effects of the neutrino background on nonadiabatic neutrino  
flavor evolution in the region above the neutrino sphere are  
potentially more significant than are the neutrino background effects  
on adiabatic neutrino flavor evolution. In general, the evaluation of  
$B$ and $B_{e\tau}$ from Eqs. (22a \& b) is considerably more  
complicated when neutrino flavor evolution is nonadiabatic than it is  
when the adiabatic limit for neutrino flavor evolution obtains.

A neutrino of energy $E_{\rm R}$, nonadiabatically going through a  
resonance at a point above the neutrino sphere, experiences a  
neutrino background effect which depends on the prior histories of  
all the neutrinos in the ensemble which are passing through the  
resonance region. In this case, we cannot argue that background  
neutrinos with $E_\nu<E_{\rm R}$ go through resonances adiabatically.  
The flavor evolution for background neutrinos with $E_\nu>E_{\rm R}$  
can still be considered adiabatic for the purposes of calculating $B$  
and $B_{e\tau}$, since these neutrinos will not yet have gone through  
resonances when they are in the resonance region for energy $E_{\rm  
R}$.

In Fig. 8 we graphically illustrate the difficulties inherent in  
computing $B$ and $B_{e\tau}$ from Eqs. (22a \& b) for nonadiabatic  
neutrino flavor evolution. In this figure we show the radial path of  
a neutrino with energy $E_{\rm R}$. The resonance position for this  
neutrino is the point labeled RES$(E_{\rm R})$. The path for a  
neutrino of energy $E_{\rm B}$ representative of the neutrino  
background at the point RES$(E_{\rm R})$ is labeled by $E_{\rm B}$.  
If $E_{\rm B}<E_{\rm R}$ then the neutrino on the path labeled by  
$E_{\rm B}$ presumably propagated through a resonance of its own  
prior to reaching position RES$(E_{\rm R})$. The resonace position  
for the background neutrino is labeled RES$(E_{\rm B})$. Whether or  
not this background neutrino experiences flavor conversion at  
RES$(E_{\rm B})$ depends, in turn, on the flavor evolution histories  
of the background neutrinos which pass through this point. The paths  
for some of these ``secondary'' background neutrinos are shown in Fig. 8.

As we can see from Fig. 8, an exact calculation of the neutrino  
background contributions requires us to simultaneously follow the  
flavor evolution histories of neutrinos with different energies on  
all possible neutrino paths above the neutrino sphere. This could be  
done in a Monte Carlo calculation. However, there is a simpler  
alternative if we make note of the following two facts. First, we are
most  interested in regions which are far away from the  
neutrino sphere. The region for $r$-process nucleosynthesis in the  
hot bubble is located at radii $r>4R_\nu$. So the polar angles for  
neutrino paths to a point in this region lie in a narrow range around  
$\theta_{\bf q}=0$. In addition, at a point close to the neutrino  
sphere where the polar angles for the relevant neutrino paths can be  
significantly different from zero, the electron number density is so  
high that neutrino background effects can be safely ignored.  
Therefore, we can make an approximation and take the flavor evolution
history of a radially propagating neutrino ($\theta_{\bf q}=0$) as
representative of the flavor evolution histories of all neutrinos with
the same energy.

The flavor evolution history of radially propagating neutrinos for a  
given set of $\delta m^2$ and $\sin^22\theta$ can then be calculated  
with the following procedure:

(1$^\prime$.) We numerically represent the neutrino energy spectrum
with a grid of energy bins. These energy bins cover a neutrino energy
range of 1--100 MeV. Typically our numerical calculations employ $\sim
200$ energy bins. Since neutrinos with lower energies go through
resonances first, we  
start the calculations at the lower end of the energy grid.

(2$^\prime$.) For the particular grid point (neutrino energy bin) at
neutrino energy $E_\nu$, we use  
the iterative procedure outlined at the beginning of this section to  
locate the resonance position, $t_{\rm res}(E_\nu)$, for this  
particular neutrino energy $E_\nu$. As a byproduct of this iterative
procedure, we will obtain the corresponding neutrino  
background contributions $B$ and $B_{e\tau}$ at this position $t_{\rm
res}(E_\nu)$. The evaluation of $B$  
and $B_{e\tau}$ in this case is quite similar to that for the case of adiabatic  
neutrino flavor evolution, except that here we must use Eqs. (22a \& b)  
together with the flavor evolution histories of neutrinos with  
energies lower than $E_\nu$.

(3$^\prime$.) Using the resonance position, $t_{\rm res}(E_\nu)$, and the  
corresponding neutrino background contributions $B$ and $B_{e\tau}$  
from Step (2$^\prime$), we can evaluate the Landau-Zener probability $P_{\rm  
LZ}(E_\nu)$ (Eq. [28]) for a neutrino with energy $E_\nu$ to jump  
from one mass eigenstate to the other in the course of transversing  
the resonance region.

(4$^\prime$.) The flavor evolution history of $\nu_e$ neutrinos with energy  
$E_\nu$ is then approximated as
$$|a_{1e}(t)|^2\approx\cases{0&if $t\le t_{\rm res}(E_\nu)$;\cr
                            P_{\rm LZ}(E_\nu)&otherwise.\cr}\eqno(39)$$
Likewise, the flavor evolution history of $\nu_\tau$ neutrinos with  
energy $E_\nu$ is approximated as
$$|a_{1\tau}(t)|^2\approx\cases{1&if $t\le t_{\rm res}(E_\nu)$;\cr
                           1-P_{\rm LZ}(E_\nu)&otherwise.\cr}\eqno(40)$$
In the above two equations, the evolutionary parameter $t$ increases  
away from the neutrino sphere. These approximations for the neutrino flavor  
evolution history, together with Eqs. (22a \& b), are then used in  
the iterative procedure in Step (2$^\prime$) to locate the resonance position  
and calculate the corresponding neutrino background contributions for  
neutrinos with energies higher than $E_\nu$.

At the end of the above procedure, we will have obtained the
approximate flavor evolution histories for all the neutrino energies on
the energy grid. This information then can be used to calculate the
electron fraction $Y_e$  
in the $r$-process nucleosynthesis region as described in Ref. [6].  
We present the new $Y_e=0.5$ line, including the neutrino background
effects, as a dotted contour line on the  
$(\delta m^2,\ \sin^22\theta)$ plot in Fig. 9. The original  
$Y_e=0.5$ line in Fig. 2 of Ref. [6] is shown as the solid contour  
line in Fig. 9. To the right of the $Y_e=0.5$ line, the material will be  
driven too proton rich for $r$-process nucleosynthesis to occur in the hot bubble.

By examining the two contour lines in Fig. 9, we can draw two  
conclusions. First, with a proper treatment of the neutrino background
effects, we see that $r$-process nucleosynthesis in  
the hot bubble remains a sensitive probe of the flavor-mixing  
properties of neutrinos with cosmologically significant masses. In fact,  
inclusion of the neutrino background contributions results in a small  
modification of the original $Y_e=0.5$ line for $\delta  
m^2=4$ eV$^2$ to $\delta m^2=10^4$ eV$^2$. Furthermore, after we take
into account the neutrino background contributions, it is evident that
the range of neutrino vacuum  
mass-squared difference $\delta m^2$ probed by $r$-process
nucleosynthesis is {\it extended} down to $\delta  
m^2< 2$ eV$^2$. The reason for this extension can be found in the
nonlinear nature of  
neutrino flavor transformation in the presence of a neutrino background.

Close to the neutrino sphere where little neutrino flavor  
transformation has occurred, the number density of $\nu_e$ neutrinos  
is larger than that of $\nu_\tau$ neutrinos. This is because the  
luminosities for $\nu_e$ and $\nu_\tau$ are approximately the same,  
but the average $\nu_\tau$ neutrino energy is much higher (cf. Eq.  
[12b]). However, with neutrino flavor transformation, more $\nu_e$  
neutrinos are transformed into $\nu_\tau$ neutrinos than $\nu_\tau$  
neutrinos are transformed into $\nu_e$ neutrinos.
This is because there are more low energy $\nu_e$ neutrinos and only  
low energy neutrinos are very efficiently transformed for the  
parameters along the dotted contour line in Fig. 9. Because of the  
nonlinear evolution of the neutrino background, the diagonal  
contribution $B$ evolves from a positive value for positions close to the neutrino  
sphere to a negative value for positions far away from the neutrino
sphere. Neutrinos with $\delta m^2<2$ eV$^2$ and energies over a broad
range will tend to have resonances far enough out that the diagonal
contributions will satisfy $B<0$. For a given $\delta m^2$ and a given
energy $E_\nu$, the resonance position will lie closer to the neutrino
sphere for the case $B<0$ than it would for the case where no neutrino
background is present (cf. Eq. [26]).

As Ref. [6] discusses, $Y_e$ and, hence, $r$-process nucleosynthesis
are sensitive to neutrino flavor conversion only when resonances occur
inside the weak freeze-out radius. The weak freeze-out radius is the
radius beyond which typical $\nu_e$ and $\bar\nu_e$ capture rates are
small compared to the material expansion rate. When $B<0$, the
resonances for given $\delta m^2$ are drawn in toward the neutrino
sphere. Hence, we find that the $Y_e=0.5$ line drops to lower values of
$\delta m^2$ in the presence of a neutrino background.
\vskip .2in
\centerline{\bf IV. Conclusions}
\vskip .2in
We have calculated neutrino flavor transformation in the region above  
the neutrino sphere in Type II supernovae including all contributions  
from the neutrino background. In particular, we have examined the  
neutrino background effects on both cases of adiabatic and  
nonadiabatic neutrino flavor evolution. In the case of adiabatic  
neutrino flavor evolution, which is most relevant for supernova shock  
re-heating, we find that the neutrino background has a completely  
negligible effect on the range of vacuum mass-squared difference,
$\delta m^2$, and vacuum mixing angle, $\theta$, or equivalently  
$\sin^22\theta$, required for enhanced shock heating. In  
the case of nonadiabatic neutrino flavor evolution relevant for  
$r$-process nucleosynthesis in the hot bubble, we find that  
$r$-process nucleosynthesis from neutrino-heated supernova ejecta  
remains a sensitive probe of the mixing between a light $\nu_e$ and a  
$\nu_{\tau(\mu)}$ with a cosmologically significant mass. The  
modification of the $(\delta m^2,\ \sin^22\theta)$ parameter region  
probed by $r$-process nucleosynthesis due to the neutrino background  
effects is generally small. The nonlinear nature of neutrino  
flavor transformation in the presence of a neutrino background actually 
{\it extends} the sensitivity of $r$-process nucleosynthesis to smaller  
values of $\delta m^2$.

In general, we find that a proper account of neutrino background
effects leads to no modification in the overall qualitative conclusions
of Refs. [5] and [6]. At the early epochs of the post-core-bounce
supernova environment ($t_{\rm PB}<1$ s), we find that the
characteristically large electron number densities and large density
scale heights determine the phenomenon of neutrino flavor
transformation. Even at the later epochs associated with $r$-process
nucleosynthesis, the effects of the neutrino background on neutrino
flavor evolution are small.
\vskip .2in
\centerline{\bf Acknowledgments}
\vskip .2in
We want to thank J. R. Wilson and R. W. Mayle for much patient
education on the subject of supernova neutrinos. We would like to
acknowledge discussions with A. B. Balantekin and W. C. Haxton. Y.-Z.
Qian also acknowledges discussions with M. Herrmann and M. Burkardt.
This work was supported by the Department of Energy under Grant No.
DE-FG06-90ER40561 at the Institute for Nuclear Theory and by NSF Grant
No. PHY-9121623 at UCSD.
\vfill
\eject
\centerline{\bf References}
\vskip .2in
\noindent[1] G. Sigl and G. Raffelt, Nucl. Phys. {\bf B406}, 423  
(1993).

\noindent[2] S. Samuel, Phys. Rev. D {\bf 48}, 1462 (1993).

\noindent[3] G. M. Fuller, R. W. Mayle, J. R. Wilson, and D. N.
Schramm, Astrophys. J. {\bf 322}, 795 (1987).  

\noindent[4] D. N\"otzold and G. Raffelt, Nucl. Phys. {\bf B307}, 924 (1988).
 
\noindent[5] G. M. Fuller, R. W. Mayle, B. S. Meyer, and J. R.  
Wilson, Astrophys. J. {\bf 389}, 517 (1992).

\noindent[6] Y.-Z. Qian, G. M. Fuller, G. J. Mathews, R. W. Mayle, 
J. R. Wilson, and S. E. Woosley, Phys. Rev. Lett. {\bf71}, 1965  
(1993).

\noindent[7] L. Wolfenstein, Phys. Rev. D {\bf 17}, 2369 (1978); {\bf  
20}, 2634 (1979); S. P. Mikheyev and A. Yu. Smirnov, Nuovo Cimento  
Soc. Ital. Fis. {\bf 9C}, 17 (1986); H. A. Bethe, Phys. Rev. Lett.  
{\bf 56}, 1305 (1986).

\noindent[8] W. C. Haxton, Phys. Rev. D {\bf 36}, 2283 (1987).
\vfill
\eject
\centerline{\bf Figure Captions}
\vskip .2in
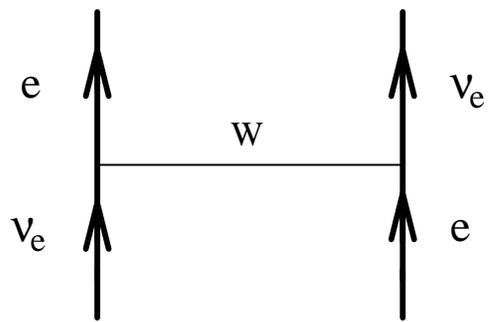
\noindent{\bf Fig. 1} A generic Feynman graph for $\nu_e$-$e$ scattering.

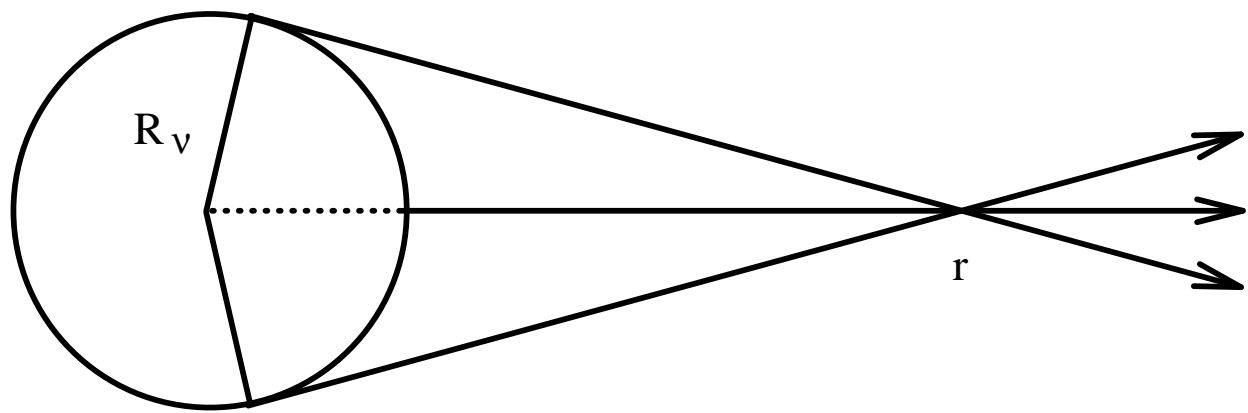
\noindent{\bf Fig. 2} The geometrical arrangement of a neutrino sphere  
with radius $R_\nu$, a point above the neutrino sphere at radius $r$,  
and various neutrino paths.

\noindent{\bf Fig. 3} Generic Feynman graphs for neutrino-neutrino  
exchange-scattering processes. Fig. 3a is for $\nu_e$-$\nu_e$ scattering and  
Fig. 3b is for $\nu_\tau$-$\nu_\tau$ scattering.

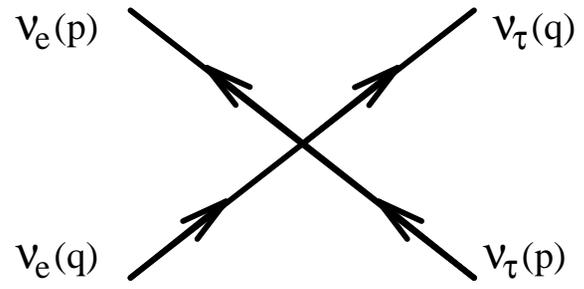
\noindent{\bf Fig. 4} Graphic representation for off-diagonal  
contributions from the neutrino background.

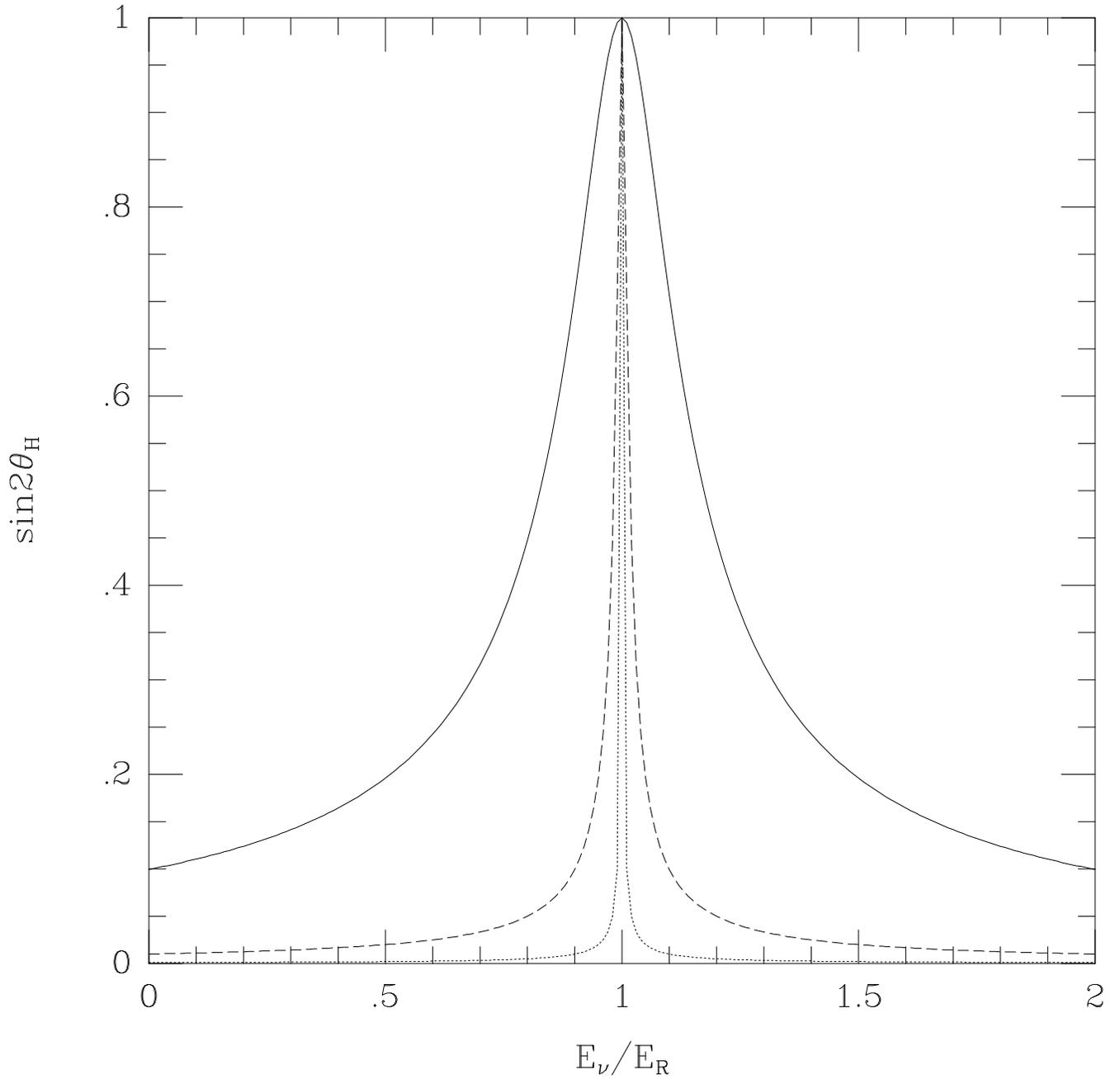
\noindent{\bf Fig. 5} The zero-order expression for $\sin2\theta_{\rm H}$  
as a function of $E_\nu/E_{\rm R}$ for three different vacuum mixing  
angles. The dotted line corresponds to $\tan2\theta=10^{-3}$. The  
dashed line corresponds to $\tan2\theta=10^{-2}$, while the solid  
line is for $\tan2\theta=0.1$.

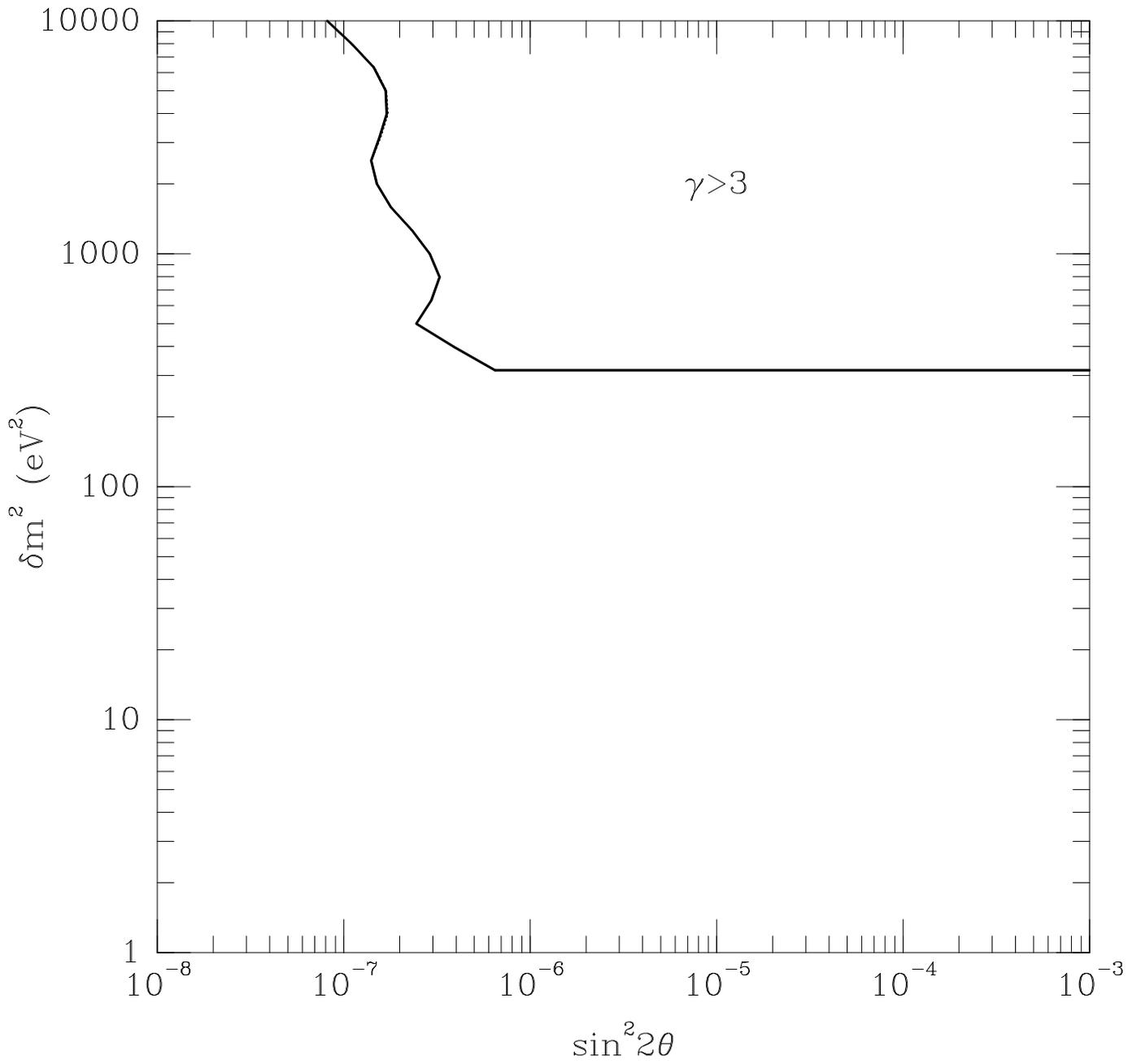
\noindent{\bf Fig. 6} Contour lines for $\gamma=3$ on the ($\delta m^2,\  
\sin^22\theta$) plot for the shock re-heating epoch.
The solid contour line is calculated for the bare electron number  
density. The dotted line, which cannot be distinguished from the  
solid line in this case, is calculated with the neutrino background  
contributions.

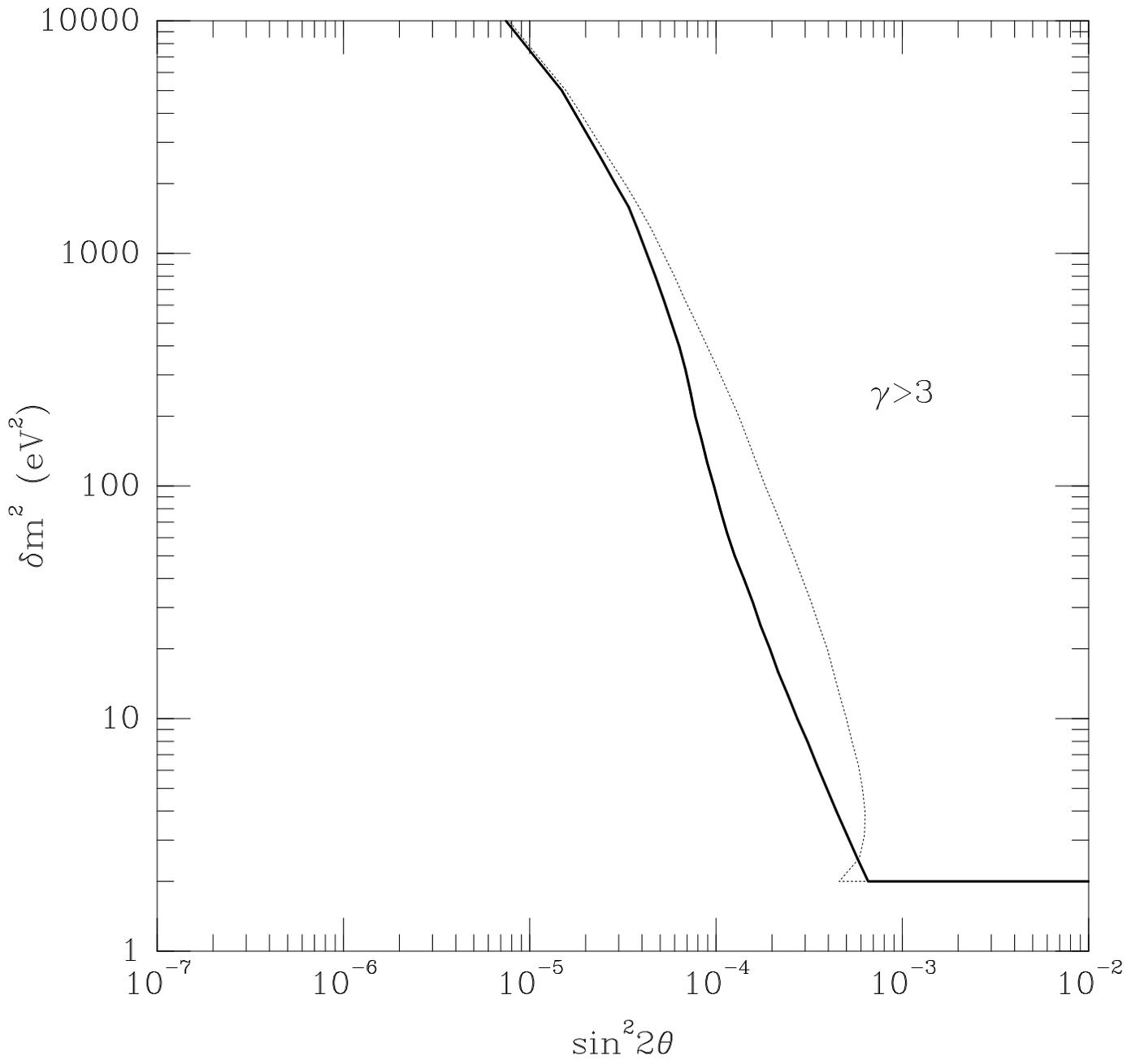
\noindent{\bf Fig. 7} As in Fig. 6, but for the hot bubble/$r$-process  
nucleosynthesis epoch.

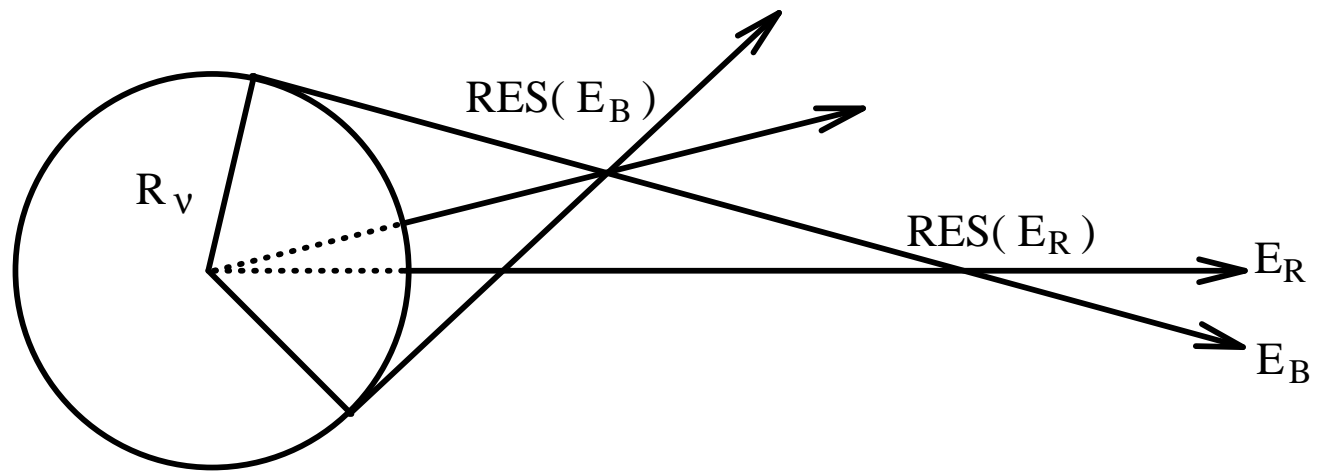
\noindent{\bf Fig. 8} Illustration of the difficulties inherent in  
computing the neutrino background contributions $B$ and $B_{e\tau}$  
for the case of nonadiabatic neutrino flavor evolution. The radial path
of a neutrino with energy $E_{\rm R}$ and resonance position
RES$(E_{\rm R})$ is shown. The path for a neutrino of energy $E_{\rm
B}$ representative of the neutrino background at position RES$(E_{\rm
R})$ is shown together with its resonance position RES$(E_{\rm B})$.
Paths for background neutrinos at position RES$(E_{\rm B})$ are also shown.

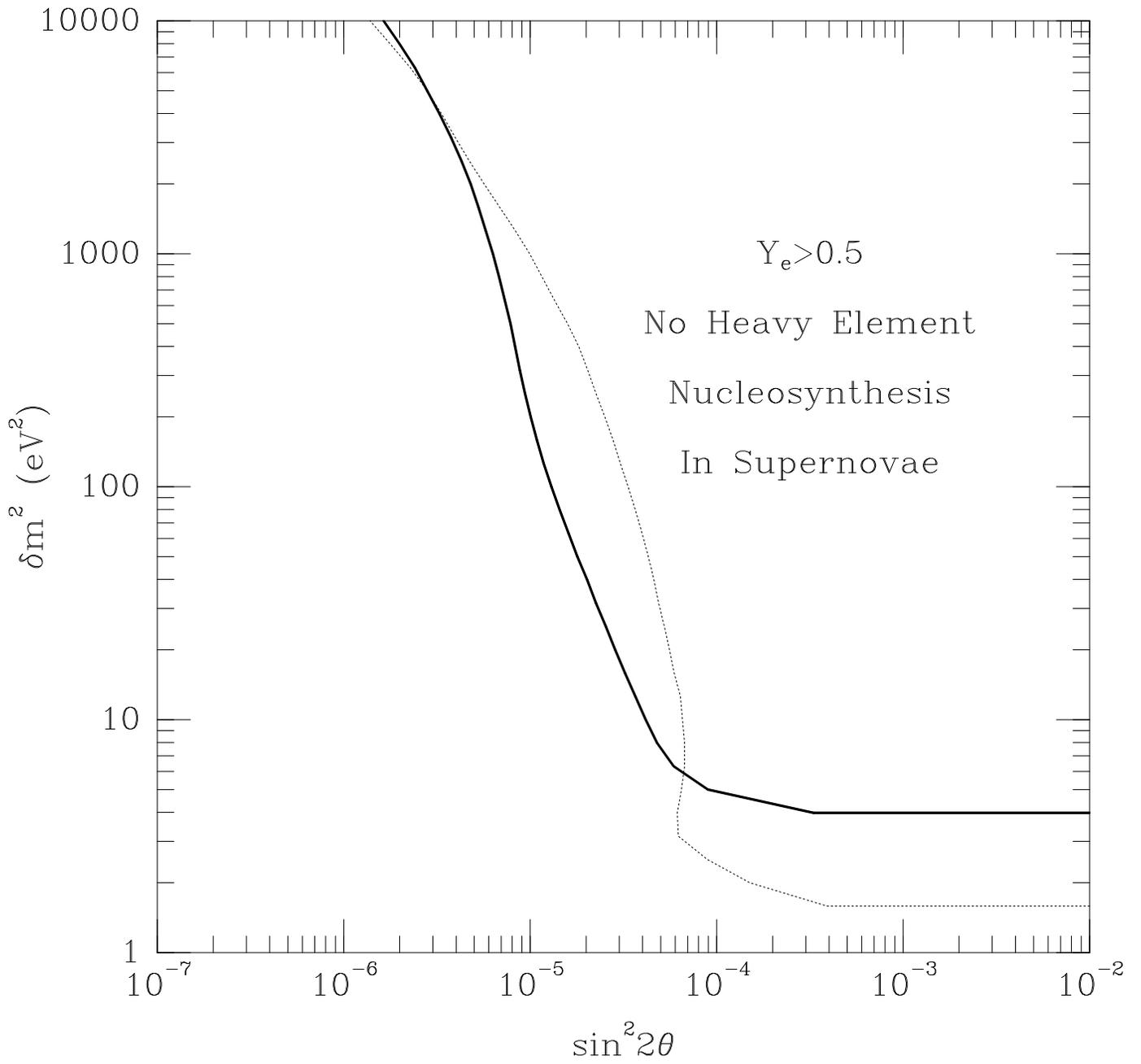
\noindent{\bf Fig. 9} Contour lines for $Y_e=0.5$ are shown on the
($\delta m^2,\ \sin^22\theta$) plot. The solid line is the same as the
$Y_e=0.5$ line in Fig. 2 of Ref. [6], whereas the dotted line is calculated with  
the full neutrino background contributions.
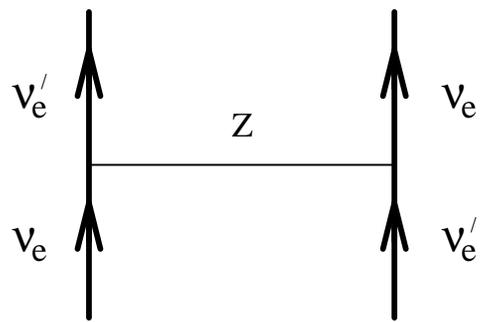
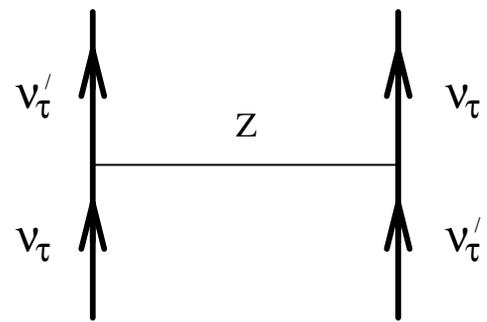
\end